\documentclass[sn-nature]{sn-jnl}% Style for submissions to Nature Portfolio journals
\usepackage{graphicx}%
\usepackage{multirow}%
\usepackage{amsmath,amssymb,amsfonts}%
\usepackage{amsthm}%
\usepackage{mathrsfs}%
\usepackage[title]{appendix}%
\usepackage{xcolor}%
\usepackage{textcomp}%
\usepackage{manyfoot}%
\usepackage{booktabs}%
\usepackage{algorithm}%
\usepackage{algorithmicx}%
\usepackage{algpseudocode}%
\usepackage{listings}%
\usepackage{color}
\usepackage{soul}
\defcitealias{Garcia-Aspeitia2018}{GA2018}

\definecolor{v}{rgb}{0.6, 0.2, 0.8} %comentarios VM

\theoremstyle{thmstyleone}%
\theoremstyle{thmstyletwo}%

\theoremstyle{thmstylethree}%

\begin{document}

\title[Article Title]{Synchronize your \textit{chrono-brane}: Testing a variable brane tension model with strong gravitational lensing}

%%=============================================================%%
%% Prefix	-> \pfx{Dr}
%% GivenName	-> \fnm{Joergen W.}
%% Particle	-> \spfx{van der} -> surname prefix
%% FamilyName	-> \sur{Ploeg}
%% Suffix	-> \sfx{IV}
%% NatureName	-> \tanm{Poet Laureate} -> Title after name
%% Degrees	-> \dgr{MSc, PhD}
%% \author*[1,2]{\pfx{Dr} \fnm{Joergen W.} \spfx{van der} \sur{Ploeg} \sfx{IV} \tanm{Poet Laureate} 
%%                 \dgr{MSc, PhD}}\email{iauthor@gmail.com}
%%=============================================================%%

\author*[1]{\fnm{Tom\'as } \sur{Verdugo}}\email{tomasv@astro.unam.mx}

\author[2]{\fnm{Mario H.} \sur{Amante}}\email{mario.herrera@fisica.uaz.edu.mx}
%\equalcont{These authors contributed equally to this work.}

\author[3]{\fnm{Juan} \sur{Maga\~na}}\email{juan.magana@ucentral.cl}
%\equalcont{These authors contributed equally to this work.}

\author[4]{\fnm{Miguel A.} \sur{Garc\'{\i}a-Aspeitia}}\email{angel.garcia@ibero.mx}

\author[5]{\fnm{Alberto} \sur{Hern\'andez-Almada}}\email{ahalmada@uaq.mx}

\author[6]{\fnm{Ver\'onica} \sur{Motta}}\email{veronica.motta@uv.cl}

\affil*[1]{\orgname{Instituto de Astronom\'ia, Observatorio Astron\'omico Nacional, Universidad Nacional Aut\'onoma de M\'exico}, \orgaddress{\street{Apartado postal 106, C.P. 22800}, \city{Ensenada}, \country{M\'exico}}}

\affil[2]{\orgdiv{} \orgname{Unidad Acad\'emica de F\'isica, Universidad Aut\'onoma de Zacatecas}, \orgaddress{\street{Calzada Solidaridad esquina con Paseo a la Bufa S/Nt}, \city{} \postcode{C.P. 98060}, \state{Zacatecas}, \country{M\'exico}}}

\affil[3]{\orgdiv{} \orgname{Escuela de Ingenier\'ia, Universidad Central de Chile}, \orgaddress{\street{Avenida Francisco de Aguirre 0405, 171-0164}, \city{La Serena} \postcode{} \state{Coquimbo}, \country{Chile}}}

\affil[4]{\orgdiv{} \orgname{Depto. de Física y Matemáticas, Universidad Iberoamericana Ciudad de México}, \orgaddress{\street{Prolongación Paseo de la Reforma 880, 01219}, \city{} \postcode{} \state{M\'exico},\country{M\'exico}}}

\affil[5]{\orgdiv{} \orgname{Facultad de Ingenier\'ia, Universidad Aut\'onoma de Quer\'etaro}, \orgaddress{\street{Centro Universitario Cerro de las Campanas}, \city{} \postcode{76010} \state{Santiago de Quer\'etaro},\country{M\'exico}}}

\affil[6]{\orgdiv{} \orgname{Instituto de F\'isica y Astronom\'ia, Universidad de Valpara\'iso}, \orgaddress{\street{Avda. Gran Breta\~na, 1111}, \city{} \postcode{} \state{Valpara\'iso},\country{Chile}}}

%This is the one%
%This is the best choice!
%%==================================%%
%% sample for unstructured abstract %%
%%==================================%%

\abstract{Brane world models have shown to be promising  to understand the late cosmic acceleration, in particular because such acceleration can be naturally derived, mimicking the dark energy behaviour just with a five dimensional geometry. In this paper we present a strong lensing joint analysis using a compilation of early-type galaxies acting as a lenses, united with the power of the well studied strong lensing galaxy cluster Abell\,1689. We use the strong lensing constraints to investigate a brane model with variable brane tension as a function of the redshift. In our joint analysis we found a value $n = 7.8^{+0.9}_{-0.5}$, for the exponent related to the brane tension,  showing that $n$ deviates from a Cosmological Constant (CC) scenario (n=6). We obtain a value for the  deceleration parameter, $q(z)$ today, $q(0)=-1.2^{+0.6}_{-0.8}$, and a transition redshift, $z_t=0.60\pm0.06$ (when the Universe change from an decelerated phase to an accelerated one). These results are in contrast with previous work that favors CC scenario, nevertheless our lensing analysis is in agreement with a formerly reported conclusion  suggesting that the variable brane tension model is able to source a late cosmic acceleration without an extra fluid as in the standard one.}

%%================================%%
%% Sample for structured abstract %%
%%================================%%


%%\pacs[JEL Classification]{D8, H51}

%%\pacs[MSC Classification]{35A01, 65L10, 65L12, 65L20, 65L70}

\maketitle

\section{Introduction}

The accelerated expansion of the universe is one of the most profound conundrums in modern cosmology, being firstly evidenced by Type Ia Supernovae \citep[SnIa][]{Riess,Perlmutter1999} and confirmed independently by measurements of the Cosmic Microwave Background Radiation anisotropies \citep[CMB,][]{2001PhRvL..86.3475J,Pryke2002}, among other observations \citep{Weinberg2013}. Under the general relativity theory, the cosmological constant (CC), arising from quantum vacuum fluctuations, is the simplest explanation as source of this acceleration because it is consistent with several cosmological tests \citep{Carroll2001,Bull2016,Cyburt2016}. Nevertheless, there is an inconsistency ($\sim 120$ orders in magnitude) between the theoretical estimation and astrophysical observations, \citep[see][]{Copeland:2006wr}. Besides, this CC hypothesis does not offer an explanation of why the Universe is accelerating today ($z\sim0.7$) but not in another  cosmic epoch \citep[see][]{Copeland:2006wr,Bamba2012}.

Braneworld models are plausible alternatives to the problem of the accelerated expansion of the universe \citep{Randall-I,Randall-II,Dvali:2000hr,Gergely,Gergely:2008jr,Sahni:2002dx,Sahni:2005pf,Abdalla:2009,Aros:2016wpv}. In these models, our four dimensional observable Universe is a brane embedded in a five dimensional bulk and the cosmic acceleration has a geometric nature. Some authors \citep{Garcia-Aspeitia:2016kak} investigated a dark energy (DE) component in a braneworld geometry with constant brane tension using several cosmological data. They found a tension on the constraints obtained from low and high redshift data which suggests that the evidence for the existence of a brane is not significant.
Later on, in a further study \citep{Garcia-Aspeitia2018}(hereafter  \citetalias{Garcia-Aspeitia2018}), it was  shown that it is possible to drive the late-time cosmic acceleration with a five-dimensional geometry without a DE component by considering a variable brane tension (hereinafter, VBT) modeled with a polynomial function; the authors called it the \textit{chrono-brane} model. By combining several cosmological data, SNIa, CMB, baryon acoustic oscillations (BAO), and observational Hubble data (OHD) and their joint analysis,  they constrained the exponent of the VBT function as $n = 6.19 \pm 0.12$ which provides an extra term in the Friedmman equation acting as a CC component at cosmic late times. Nevertheless, the VBT model should be investigated at both galactic and galaxy cluster scales. The last are the largest structures in the Universe and they provide insights into the transition of the linear and non linear cosmological perturbations, i.e. in the late dynamics of the Universe. Thus, it is necessary to consider more cosmological data to shed light on 
the properties of the VBT model at these scales.

In this work we extend that study using observations from strong gravitational lensing at galaxy and galaxy clusters system scales, which in the last decade have been considered more frequently in  cosmological parameter estimation \citep[see][and references therein]{Cao2015,Magana2018a,Caminha2022,Shajib2022}. In particular cluster strong lensing cosmography has been used in the past in a successfully way, but it requires high-quality data, for example spectroscopy to measure redshifts of multiple images. However, using a galaxy cluster to constraint cosmological models has his caveats, many systematic effects can affect an individual strong-lensing model, for example intrinsic degeneracies in the models, mass components external to the cluster, or line-of-sight perturbers \citep{Caminha2022}. As comment by these authors, a possible solution is to perform a combined cosmographical analysis using several clusters. In the present work we propose a different approach which consist in combining two lensing constraints at different scales: the power of the well known lensing galaxy cluster Abell 1689 \citep{Jullo:2010}, and the capability of  early-type galaxies acting as lenses \citep{Mario:2020}.

This paper is organized as follows. In Section \ref{Sec2}, we present the theoretical model. In Section \ref{Sec3}, we summarize our data and methodology. In Section \ref{Sec4}, we describe our results, and we discuss them in Section \ref{Sec5}. Finally, we present our conclusions  and remarks in Section \ref{Sec6}.

%%%%%%%%%%%%%%%%
\begin{figure}
\includegraphics[width=\columnwidth]{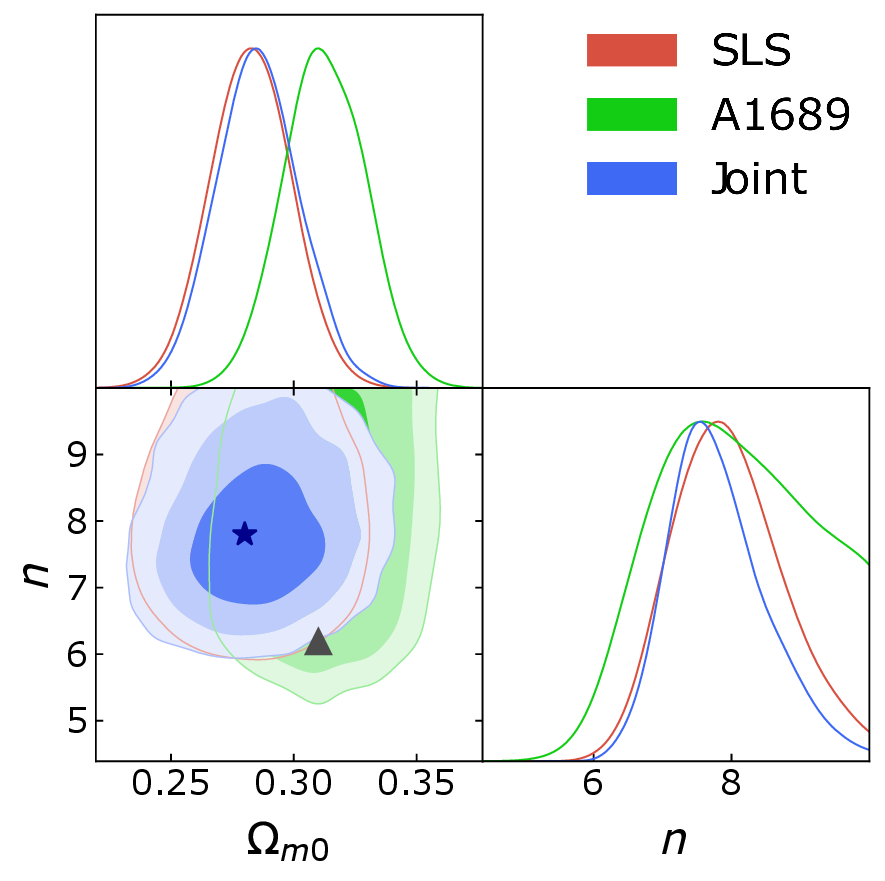}
\caption{1D marginalized posterior distributions and the 2D confidence contours ($1\sigma, 2\sigma, 3\sigma$) for the $\Omega_{m0}$ and $n$ parameters of the VBT model for early-type galaxies (red) and galaxy cluster Abell 1689 (dark-blue) assuming a  Gaussian prior on $\Omega_{m0}$ parameter. The star indicates the mean value for the joint, and the triangle the mean value reported by \citetalias{Garcia-Aspeitia2018}. \label{fig:BV1}}
\end{figure}
%%%%%%%%%%%%%%%%%%%

%%%%%%%%%%%%%%%%%%%%%%%%%%%%%%%%%%%%%%%%%%%%%%%%%%%%%%%%%%%%%%%%%%
\section{Theoretical model} \label{Sec2}
%%%%%%%%%%%%%%%%%%%%%%%%%%%%%%%%%%%%%%%%%%%%%%%%%%%%%%%%%%%%%%%%%%

 The theoretical model that we are going to constrain,  namely the  VBT model or \textit{chrono-brane} model  \citepalias{Garcia-Aspeitia2018}, is a function of the scale factor (redshift), i.e. $\lambda$(a) or $\lambda$(z). In particular, the authors propose a polynomial function for the brane tension which is dominant in later times in the Universe evolution, but subdominant in the early Universe. We refer the reader to such work for a rigorous description, here we discuss briefly the theoretical framework. We start with the VBT field equation as
\begin{equation}
    G_{\mu\nu}-8\pi GT_{\mu\nu}=\frac{1}{\lambda}\left[48\pi G\Pi_{\mu\nu}+\frac{3}{4\pi G}\xi_{\mu\nu}\right], \label{fe}
\end{equation}
with
\begin{eqnarray}
&&\xi_{\mu\nu}=\mathcal{U}\left(u_{\mu}u_{\nu}+\frac{1}{3}\epsilon_{\mu\nu}\right)+\mathcal{P}_{\mu\nu}, \\
&&\Pi_{\mu\nu}=-\frac{1}{4}T_{\mu\alpha}T^{\alpha}_{\nu}+\frac{1}{12}T^{\alpha}_{\alpha}T_{\mu\nu}+\frac{1}{24}g_{\mu\nu}[3T_{\alpha\beta}T^{\alpha\beta}-(T^{\alpha}_{\alpha})^2].
\end{eqnarray}
where $G_{\mu\nu}$ is the standard Einstein tensor, $G$ is the Newton gravitational constant, $\lambda$ is the brane tension, $\xi_{\mu\nu}$ is a non-local Weyl tensor decomposed in its  irreducible form which also contains $\mathcal{U}$ as the non-local energy density, $\mathcal{P}_{\mu\nu}$ is the non-local anisotropic stress tensor, $u_{\mu}$ is the four-velocity and $\epsilon_{\mu\nu}\equiv g_{\mu\nu}+u_{\mu}u_{\nu}$, being $g_{\mu\nu}$ the metric tensor associated. In addition, $T_{\mu\nu}$ and $T=T^{\alpha}_{\alpha}$ are the standard energy-momentum tensor and scalar respectively; finally $\Pi_{\mu\nu}$ contains a quadratic form of the energy-momentum tensor.

 The corrective term that come from brane world is related to the brane tension that is given by $\lambda$, which in this model is not a constant. Therefore, the low energy limit is considered when $\lambda\to\infty$, where the traditional field equation of General Relativity is recovered, and for $\lambda\to0$ the extra term play a preponderant role. Finally, notice that in this case we do not consider extra fields onto the bulk, neglecting those terms that come from the non local tensor and only taking those fields that live in the brane.

To study the brane cosmological dynamics, we consider the standard homogeneous and isotropic line element of Friedmann-Lema\^itre-Robertson-Walker (FLRW) for a flat geometry,

\begin{equation}
ds^2=-dt^2+a(t)^2(dr^2+r^2d\Omega^2),
\end{equation}
where $d\Omega^2\equiv d\theta^2+\sin^2\theta d\varphi^2$ is the solid angle in spherical coordinates and $a(t)$ is the scale factor. Moreover, the energy-momentum tensor will be written as
\begin{equation}
    T_{\mu\nu}=pg_{\mu\nu}+(\rho+p)u_{\mu}u_{\nu}, \label{emt}
\end{equation}
where $p$ and $\rho$ are the pressure and density of the fluid respectively. Therefore, if we introduce the previously defined line element in Eq. \eqref{fe} together with the perfect fluid energy-momentum tensor (Eq. \eqref{emt}),  in the VBT model the Friedmann equation can be written in terms of the redshift as
\begin{equation}
E(z)^2  = \Omega_{0m}(z+1)^3+\Omega_{0r}(z+1)^4+\frac{\mathcal{M}}{(z+1)^{n}}[\Omega_{0m}^2(z+1)^6+\Omega_{0r}^2(z+1)^8], \label{Friedmann}
\end{equation}
where $E(z)\equiv H(z)/H_0$, $H_0$ is the Hubble constant, $\Omega_{m0}$ is the density parameter for matter (baryons plus dark matter), $\Omega_{0r}$ is the density parameter for radiation, and $\mathcal{M}\equiv3H_0^2/16\pi G\lambda_0$. The free parameters, related with the brane tension, $\lambda_0$ and $n$ are  coupled through the equation $\lambda(z)=\lambda_{0}(z+1)^n$ which is the polynomial function proposed in \citetalias{Garcia-Aspeitia2018}, being  $z$ the redshift.

Since $\Omega_{0r}\sim 10^{-5}$, the radiation terms in Eq. \eqref{Friedmann} can be neglected when it does not dominate the Universe dynamics, and the Friedmann equation reads as

\begin{eqnarray}
E(z)^2=\Omega_{m0}(z+1)^3+
\mathcal{M}\Omega_{
m0}^2(z+1)^{6-n}.
\label{Friedmann_VBTA}
\end{eqnarray}

An interesting solution is obtained when $n=6$, because the second term in Eq. (\ref{Friedmann_VBTA})  will be a constant given by the term $\mathcal{M}\Omega_{m0}^2$, acting as a cosmological constant. Thus, the interpretation of the cosmological constant will be in the five dimensional context.
Additionally to this, we have the following expression: $\mathcal{M}=(1-\Omega_{m0})\Omega_{m0}^{-2}$,
assuming $z=0$ and the flatness condition.

Moreover, the expression for the deceleration parameter reads
\begin{equation}
    q(z)=\frac{\Omega_{m0}}{2E^2(z)}\Big\lbrace(z+1)^3+(4-n)\mathcal{M}\Omega_{m0}(z+1)^{6-n}\Big\rbrace,
\end{equation}
where $E(z)$ is given by Eq. (\ref{Friedmann_VBTA}).

%%%%%%%%%%%%%
\begin{table*}
\centering
\caption{Mean values for the model parameters, $\Omega_{m0}$ and $n$, derived from each data set and the joint analysis. $\rho_c$ is the standard critical density of the Universe. \label{tab:par}}
\begin{tabular}{cccccc}
\hline
Data set & $\chi^{2}_{min}$ & $\chi^{2}_{red}$ & $\Omega_{m0}$ &  $n$ & $\lambda_{0}/\rho_c$\\
\hline
SLS &  245.8  & 1.7 & $0.28^{+0.016}_{-0.017}$  & $7.9^{+0.9}_{-0.7}$&$0.055^{+0.008}_{-0.007}$\\
Abell 1689 &  30.4  & 2.8 & $0.31\pm0.017$ & $8.0^{+1.5}_{-0.8}$&$0.070^{+0.010}_{-0.008}$\\
Joint &  282  & 1.8 & $0.28^{+0.016}_{-0.018}$ &  $7.8^{+0.9}_{-0.5}$ & $0.056^{+0.008}_{-0.007}$ \\
\hline
\end{tabular}
\end{table*}
%%%%%%%%%%%%%

%%%%%%%%%%%%%%%%%%%%%%%%%%%%%%%%%%%%%%%%%%%%%%%%%%
\section{Data and methodology}\label{Sec3}
%%%%%%%%%%%%%%%%%%%%%%%%%%%%%%%%%%%%%%%%%%%%%%%%%%

We constrain the cosmological parameters of the VBT model,  by combining for the first time  the strong lensing effect produced by  elliptical galaxies acting as a lens, and in the galaxy cluster Abell 1689. Here we present a short description of the methods detailed in the works of \cite{Jullo:2010}, \cite{Magana:2015}, and \cite{Mario:2020}, and how we combine both methodologies.

%%%%%%%%%%%%%%%%%%%%%%%%%%%%%%%%%%%%%%%%%%%%%%%%%%%%%%%%%%%
\subsection{Strong lensing in elliptical galaxies} 
%%%%%%%%%%%%%%%%%%%%%%%%%%%%%%%%%%%%%%%%%%%%%%%%%%%%%%%%%%%

We use the compilation of early-type galaxies acting as lenses presented by \cite{Mario:2020}. Their fiducial sample consists of $N_{SL}=143$ strong lensing systems (SLS), with four measured properties: spectroscopically measured stellar velocity dispersions $\sigma$, the Einstein radius $\theta_E$, the lens redshift $z_l$ and the source redshift $z_s$.

We can constrain cosmological parameters following the steps provided by \cite{Grillo:2007iv} minimizing the chi square function given as
\begin{equation}
\chi_{\mbox{Gal}}^2 = \sum_{i=1}^{N_{SL}} \frac{ \left[ D^{th}\left(z_{l}, z_{s}; \bf{\Theta_{Cos}} \right)  -D^{obs}(\theta_{E},\sigma^2)\right]^2 }{ (\delta D^{\rm{obs}})^2},
\label{eq:chisquareSL}
\end{equation}
where we define the ratio of two angular diameter distances $D \equiv D_{ls}/D_{s}$. Thus, the theoretical ratio $D^{th}$ is calculated from Eq.\,\ref{Friedmann}, using the definition of angular diameter distance\footnote{$D(z)=\frac{c}{H_0(1+z)}\int_0^z\frac{dz^{\prime}}{E(z^{\prime})}$}. On the other hand, through the observationally measured properties, we get the observed counterpart $D^{obs}$ , and $\delta D^{\rm{obs}}$  as the error propagation of the $D^{obs}$ function.  The vector $\mathbf{\Theta_{Cos}}$ is formed by the parameters $n$ and $\Omega_m$.

%%%%%%%%%%%%%%%%%%%%%%%%%%%%%%%%%%%%%%%%%%%%%%%%%%%%%%%%%%%
\subsection{Strong lensing in Abell 1689 galaxy cluster} \label{Abell}
%%%%%%%%%%%%%%%%%%%%%%%%%%%%%%%%%%%%%%%%%%%%%%%%%%%%%%%%%%%

While different galaxy clusters has been used to test cosmology \citep[e.g.,][]{Caminha2022,Limousin2022},  galaxy cluster Abell 1689 (A\,1689) at $z$=0.184, is still competitive to perform cosmological constraints, and  it is a widely studied cluster \citep[see][]{Bina2016}.
A\,1689 is modeled as a bi-modal mass distribution, with one central, dominant large-scale potential hosting the brightest cluster galaxy (BCG) at its centre \citep[see][]{Limousin:2007}. A second large-scale potential is located at the north-east of the cluster. Additionally, the model includes the galaxy scale dark matter haloes associated with 58 cluster members.

Following \cite{Jullo:2010}, to reconstruct the A1689 mass model and simultaneously  constrain the cosmological parameters of the VBT cosmology  we use the 'family ratio' which is defined  as the angular diameter distance ratios of two images from different sources. In particular, we use the same image catalog  of \cite{Jullo:2010}, which contains 28 images from $N_f$ = 12 families, all with measured spectroscopic redshifts in the range $1.15< z_{\rm s} < 4.86$. For the present work, we adopt an error $\Delta^2$ = 0.5" in the position of the images \citep[see][]{Magana2018a}, and the models are evaluated and optimized in the image plane.

In this case, the $\chi^{2}$ for a multiple image system $i$ is defined as 
\begin{equation}\label{eq:Chi2Lens}
\chi_{i}^{2} = \sum_{j=1}^{n_i}
\frac{\left| \vec{x}_{\rm obs}^j - \vec{x}^j(\mathbf{\Theta_{W}}) \right|^2}{\Delta^{2}}\;,
\end{equation}
where $n_i$ is the number of multiple images for the source $i$, $\vec{x}_{\rm obs}^j$ is the observed position corresponding to image $j$, and $\vec{x}^j(\mathbf{\Theta_{W}})$ is the position of image $j$ predicted by the current model, whose total parameters, i.e. the cosmological parameters and the cluster parameters (see below) are included in the
vector $\mathbf{\Theta_{W}}$. Thus, $\chi^2_{\mbox{Clu}} =\sum_{1}^{N_f} \chi_{i}^{2}$, is the total chi square function in this case.

%%%%%%%%%%%%%%%%%%%%%%%%%%%%%%%%%%%%%%%%%%%%%%%%%%%%%%%%%%%
\subsection{Combining data sets}
%%%%%%%%%%%%%%%%%%%%%%%%%%%%%%%%%%%%%%%%%%%%%%%%%%%%%%%%%%%

The probability distribution function of the model parameters is computed via {\sc LENSTOOL}\footnote{Publicly available at: https://git-cral.univ-lyon1.fr/lenstool/lenstool {\sc LENSTOOL}/} code. This code uses a Bayesian Monte Carlo Markov chain algorithm to search for the most likely parameters in the modeling, and it has been used in a large number of clusters studies, and characterized in 
\cite{Jullo:2007up}. For the present work we have incorporated a routine into {\sc LENSTOOL} in order to compute the likelihood for the galactic systems in combination with the reconstruction of the mass model for A1689.

Considering the Gaussian likelihood $\mathcal{L}\propto e^{-\chi^{2}_{\mathrm{Tot}}/2}$, the total $\chi^2$ reads
\begin{equation}\label{eq:Chi2Tot}
\chi_{\mathbf{Tot}}^{2} = \chi^2_{\mbox{Gal}} + \chi^2_{\mbox{Clu}},
\end{equation}
being $\chi^{2}_{\mathrm{Tot}}$ the chi-square function constructed using the two complementary approaches regarding strong lensing measurements.

In our calculations we have assumed a 3$\sigma$  Gaussian prior in the $\Omega_{m0} = 0.311 \pm 0.0056 $ parameter according to the observations from Planck \citep{Aghanim:2020}, and we fix the $h=0.7403\pm 0.0142$ parameter (due to its independence with the present method) to the value reported by \cite{Riess:2019} using a model independent approach. We also assume a uniform prior for the $n$ parameter within the  $0<n<10$ region.

In addition to the cosmological parameters, for A\,1689 we have 19 parameters to be fit:  the six parameters (the spatial coordinates $x$, $y$; the ellipticity, $e$; angle position, $\theta$; $r_{\rm core}$, and the velocity dispersion, $\sigma$) for each of the main and secondary clumps, five parameters ($x$, $y$, $r_{\rm core}$, $r_{\rm cut}$, $\sigma_0$) for  the dominant BCG in the main  clump, and two parameters ($r_{\rm cut}^{*}$, $\sigma^*_0$) for the galaxy-scale clumps.  The uniform priors in the parameters were set defining regions around the best values reported by \cite{Jullo:2010} and \cite{Limousin:2007}, however we check that the boundaries were not reached.

%%%%%%%%%%%%%%%%%%%%%%%%%%%%%%%%%%%%%%%%%%%%%%%%%%%%%%%%%%%
\section{Results}\label{Sec4}
%%%%%%%%%%%%%%%%%%%%%%%%%%%%%%%%%%%%%%%%%%%%%%%%%%%%%%%%%%%

%%%%%%%%%%%%%%%%%%%%%%
\begin{figure}
%\begin{tabular}{cc}
\includegraphics[width=\columnwidth]{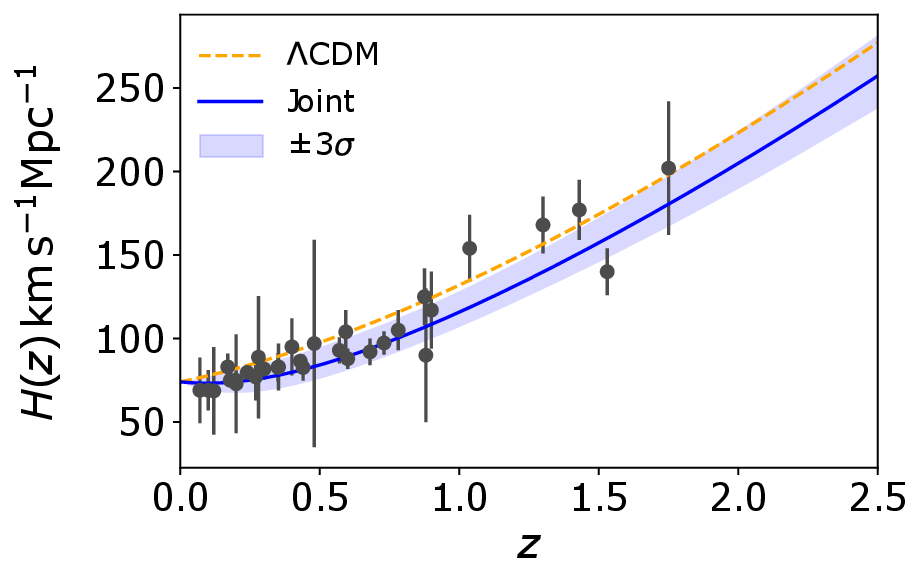}\\
\includegraphics[width=\columnwidth]{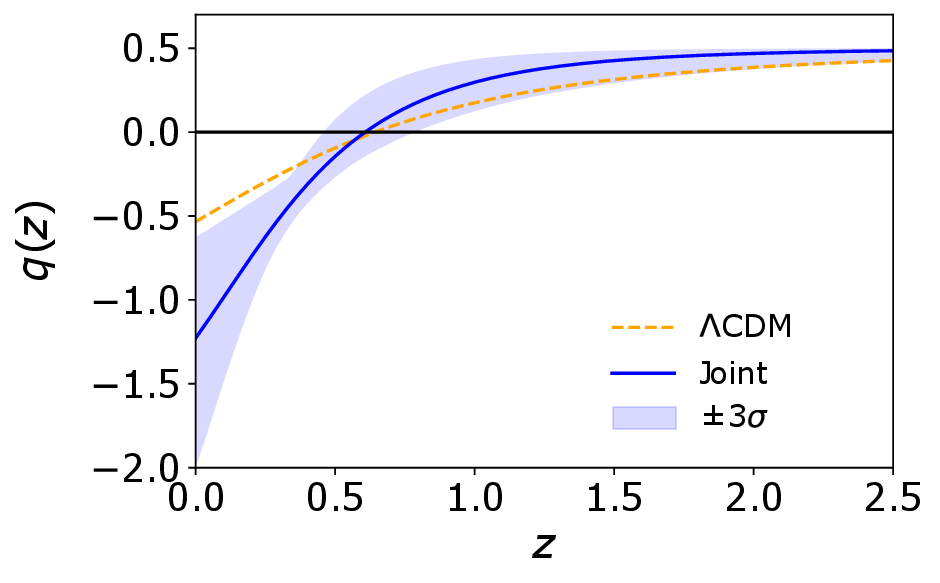}
%\end{tabular}
\caption{Comparison  between the theoretical model and $H(z)$ data (top panel) and the reconstructed $q(z)$ (bottom panel) using the results of the joint analysis constraints. The $\Lambda$CDM dynamics is plotted for comparison.}
\label{fig:qz}
\end{figure}
%%%%%%%%%%%%%%%%%%%%%%%%%

%%%%%%%%%%%%%%%%%%%%%%
\begin{figure}
\includegraphics[width=\columnwidth]{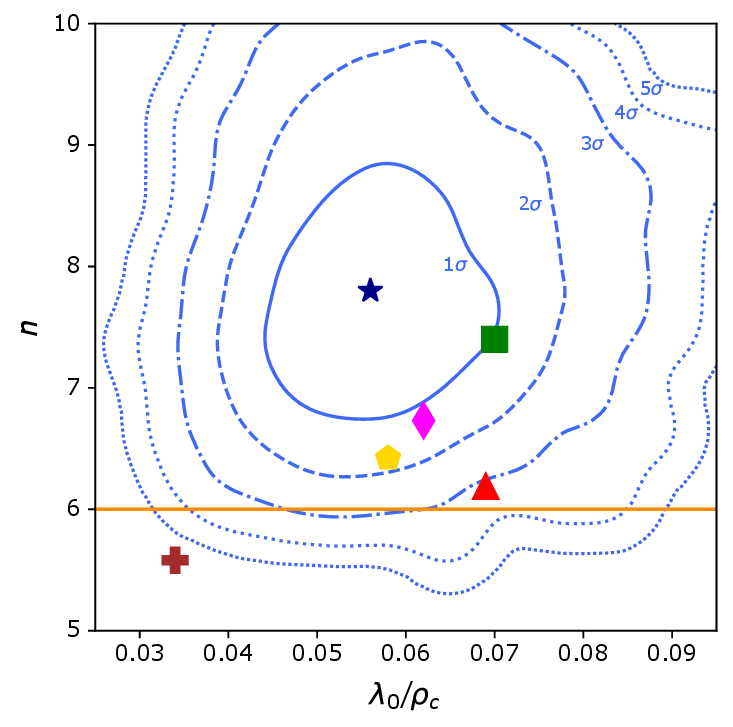}
\caption{The confidence contours ($1\sigma, 2\sigma, 3\sigma, 4\sigma$, and $5\sigma$) for the $\lambda_{0}/\rho_c$ and $n$ parameters of the VBT model from the joint analysis. The star indicates the mean value for the joint of the present work. The square, diamond, pentagon, plus, and triangle markers represent the mean values obtained from H(z), BAO, CMB, SNIa data, and the joint reported by \citetalias{Garcia-Aspeitia2018} respectively. The horizontal line represents $n=6$, i.e., a model with CC. \label{fig:n-l}}
\end{figure}
%%%%%%%%%%%%%%%%%%%%%%%%

Table \ref{tab:par}  presents the mean values for $\Omega_{m0}$ and $n$ parameters for the two data sets discussed in this work  as well as those for the joint analysis.  In the Figure \ref{fig:BV1} we show  the PDFs and the contours for these parameters. Note that strong lensing observations at different scales (galaxy cluster and galaxies) are consistent at 1$\sigma$ between both data sets. Although the result of our joint model deviates more than 3-$\sigma$  from those reported in the joint model of \citetalias{Garcia-Aspeitia2018}, this trend  of larger values for $n$ it is also  obtained with other astrophysical data sets,  e.g. with $H(z)$ or baryon acoustic oscillations (BAO)  measurements \citepalias{Garcia-Aspeitia2018}. The mean value $n = 7.8^{+0.9}_{-0.5}$ points towards a model that departs from $\Lambda$CDM, with $\lambda(z)$ = $\lambda_0 (1+z)^{7.8^{+0.9}_{-0.5}}$, we have a term in the Friedmman equation that takes the role of a DE  component at cosmic late times. Furthermore, this extra dimension VBT model with the joint constraints can explain the mass distribution in the galaxy cluster A1689 and their strong lensing features without the existence of dark energy fluid.

One of the advantages of using lensing cosmography, when compared with other standard probes, is the independence of the method with respect to the Hubble constant $H_0$. Thus, it is interesting to compare how the Hubble parameter is recovered by employing the results of our lensing fit. Using the result of the joint analysis we construct the function $H(z)$ and we compare it  with the  observational Hubble data measured using only differential age method compiled by \cite[][also see \cite{Magana:2018b}]{Moresco:2016mzx}. The comparison is depicted in the top panel of Figure \ref{fig:qz}. As before, we note that it is not consistent with a CC model. In the bottom panel of Figure \ref{fig:qz} we show the reconstruction of the deceleration parameter $q(z)$ for the joint analysis. We obtain that the Universe change from a decelerated phase to an accelerated one at $z_t=0.60\pm0.06$, and deviates from CC, with $q(0)=-1.2^{+0.6}_{-0.8}$ at $z=0$.

The Figure \ref{fig:n-l} illustrates the $\lambda_0/\rho_c - n$ confidence contours  ($1\sigma, 2\sigma, 3\sigma, 4\sigma$, and $5\sigma$) for the joint analysis; where $\rho_c$ is the critical density of the Universe. For comparison, we show the mean values obtained from $H(z)$ measurements, BAO, CMB, SNIa data, and the joint reported by \citetalias{Garcia-Aspeitia2018}.  Note the clear deviation from $\Lambda$CDM, on each  data set viewed independently. On one hand, the departure from $n=6$ in our joint analysis it is not surprising, since it is expected when using lensing as a cosmological test, both, at galaxy scale \citep{2019MNRAS.488.3745C, Mario:2020} and using a galaxy cluster \citep{Caminha2022,Limousin2022}. On the other hand, it is interesting that three mean values from independent cosmological constraints, falls inside the 2$\sigma$ region obtained from our joint analysis (see Figure \ref{fig:n-l}). By considering $\rho_c=8.070\times 10^{-11}h^2\,\text{eV}^4$, we estimate $\lambda_0\approx 2.51^{+0.38}_{-0.33}\times10^{-12}\text{eV}^4$ which is consistent with a tension compatible with the variable tension model where the brane terms does not affect the early physics, such as Nucleosynthesis among others, and produce a late time acceleration according to the observations. Our results are also consistent with the expected value for the CC energy density which is $\rho<10^{-10}$eV$^4$. In our case, the brane tension mimics the CC behavior and the value of its energy density coincide with the expected value of the CC.

%%%%%%%%%%%%%%%%%%%%%%%%%%%%%%%%%%%%%%%%%%%%%%%%%%%%%%%%%%%
\section{Discussion}\label{Sec5}
%%%%%%%%%%%%%%%%%%%%%%%%%%%%%%%%%%%%%%%%%%%%%%%%%%%%%%%%%%%

In the study by \citetalias{Garcia-Aspeitia2018}, they reported a VBT  expression with $\lambda(z) = \lambda_0 (1+z)^{6.2{\pm0.12}}$, proposing a brane that mimics dark energy dynamics and consistent with CC. In our current investigation, the strong lensing data provide $\lambda(z) = \lambda_0 (1+z)^{7.8^{+0.9}_{-0.5}}$,  suggesting  that the cosmic acceleration may be driven by a phantom dark energy-like. It is important to note that this effect stems from the dynamics of extra dimensions rather than a fluid with an equation of state (EoS)  characterized by $w<-1$. Additionally, in the present work, the observed deviation of the deceleration parameter from its expected value in the CC cosmology, at  z = 0, is noteworthy. The deceleration parameter $q(z = 0)=-1.2^{+0.6}_{-0.8}$, and signals the potential influence of a phantom dark energy-like component which contributes to an acceleration of the Universe beyond what is conventionally anticipated in the CC model.

To look deeper into the preceding paragraph, we can write the Friedmann equation under our constraints,

\begin{equation}
    E(z)^2=\Omega_{m0}(z+1)^3+\mathcal{M}\Omega_{m0}(z+1)^{-1.8}, \label{Efinal}
\end{equation}
thus, it is possible to notice that in the future, $z\to-1$, $E(z)^2\to\infty$ (Big Rip), due to the second term in the right side of Eq. \eqref{Efinal}, suggesting the presence of a phantom DE. In this context, it is important to emphasize that observations do not rule out phantom models in this context \citep{Aghanim:2020, 2002PhLB..545...23C, 2016EPJC...76..631B}.

Moreover, we can write the effective EoS from \citepalias{Garcia-Aspeitia2018} as

\begin{equation}
    w_{eff}(z) =  \frac{ 2q-1}{3[1 + 2\Omega_{m0}\mathcal{M} (z+1)^{3-n} ]} + \frac{ \Omega_{m0}\mathcal{M} (z+1)^{3-n}[2q-(4-n)]}{3[1 + 2\Omega_{m0}\mathcal{M} (z+1)^{3-n} ]},
\end{equation}
having for $z=0$ $w_{eff}\simeq0.013$ a dust fluid, however the acceleration is maintained because the condition
\begin{equation}
    w_{acc}<-\frac{1+\Omega_{mo}\mathcal{M}(4-n)}{3[1+2\Omega_{m0}\mathcal{M}]},
\end{equation}
is satisfied, since the EoS for acceleration demands $w_{acc}<0.47$ at $z=0$. 

When we compare our $n$ constraints from the joint analysis of strong lensing features with the previous one by \citetalias{Garcia-Aspeitia2018}, we have a tension of $\Delta n \approx 2.25 \sigma$ that can be appreciated in Figure \ref{fig:n-l}. Note that this can be written in terms of the parameter $\lambda_0$  as $\Delta\lambda_0 \approx 0.8 \sigma$ and hence a tension of $\sim 3\sigma$ in the $\lambda_0/\rho_c$-$n$ parameter space (see Fig. \ref{fig:n-l}). For the $\Omega_{m0}$ parameter, the tension between the two works is $\Delta \Omega_{m0}\approx 1.59\sigma$. 

\begin{figure}
\includegraphics[width=\columnwidth]{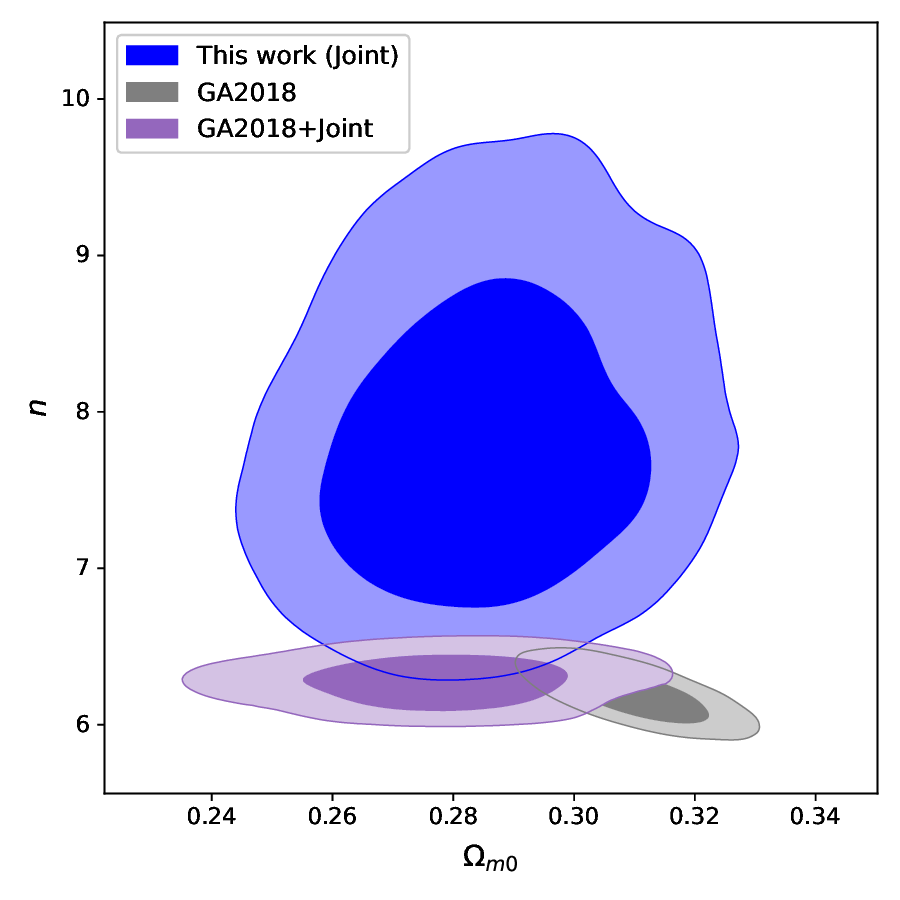}
\caption{ Comparison of confidence contours ($1\sigma$, and $2\sigma$) for $\Omega_{m0}$ and $n$ parameters. Blue contours from the Joint analysis of this work. Gray contours from the joint analysis by \citetalias{Garcia-Aspeitia2018}. The purple contours represents the joint analysis of the joints (see text). \label{fig:n-l2}}
\end{figure}

To highlight the differences among the joint constraints, we calculated a combined set of constraints on the parameters $n$ and $\Omega_{m0}$. This was achieved by multiplying the probability distribution functions of the two works under consideration: the joint analysis conducted by \citetalias{Garcia-Aspeitia2018} and our own joint analysis presented in this paper (see Figure \ref{fig:n-l2}). Note that there is a slight tension at the $2\sigma$ confidence level. The disparities in the VBT ($\Omega_{m0},n$) parameters from both joint analyses could be attributed to various factors, such as errors in the Einstein radius in the SLS, image-positional errors in the strong lensing features in A1689, among others \citep[see][]{Magana2018a,Mario:2020}. It is also worth noting that the joint constraints by \citetalias{Garcia-Aspeitia2018} are influenced by CMB data, and the conflict between early and late data emerges when we examine the data individually. To shed some light on this, we calculate the Figure-of-Merit (FoM, see \cite{Magana2018a} and references therein). This tool allows us to assess the quality of constraints from different datasets and is calculated as
\begin{equation}
\mathrm{FoM}= \frac{1}{ \sqrt{\det \mathrm{cov}(p_1,p_2,p_3,...)}}   
\end{equation}
where $\mathrm{cov}(p_1,p_2,p_3,...)$ is the covariance matrix of the $p_i$ parameters. Larger FoM values imply stronger bounds on the parameters, as they correspond to smaller error ellipses. By choosing the $\Omega_{m0}$ and $n$ parameters, we obtain the following FoM values: $24.30, 29.92,130.02,672.22,75.91,56.29$ for 
OHD, SNIa, BAO, CMB, SLS, and A1689 data, respectively. Note that the constraints of this work are stronger than those obtained from OHD and SNIa but weaker than those from BAO and CMB data.

Finally, it is important to bear in mind that the technique presented in this work is complementary to that of \citetalias{Garcia-Aspeitia2018}. This approach could assist in breaking the degeneracy between VBT parameters, obtaining stronger constraints, and determining whether this model could account for the late cosmic acceleration.

%%%%%%%%%%%%%%%
%%%%%%%%%%%%%%%%

%%%%%%%%%%%%%%%%%%%%%%%%%%%%%%%%%%%%%%%%%%%%%%%%%%%%%%%%%%%
\section{Conclusions and Remarks} \label{Sec6}
%%%%%%%%%%%%%%%%%%%%%%%%%%%%%%%%%%%%%%%%%%%%%%%%%%%%%%%%%%%

In this paper we propose a new approach combining for the first time two lensing data sets at different scales: the constraints provided by A1689 galaxy cluster and those that  come from  early-type galaxies acting as lenses. In order to test the capability of the method we study the VBT (\textit{chrono-brane}) model. This model is capable of reproducing a late cosmic acceleration in a five dimensional scenario, without assuming an extra fluid as in the standard  one, and has been study previously using different data sets. However, in the present work, we use a different methodology which is independent of $H_0$, avoiding the problem of $H_0$ tension presented in other measurements \citep{Ries2016,Riess:2019}.

We found that the joint strong lensing analysis  provides, within the errors, similar values of the parameters related to the brane tension, as those reported previously by \citetalias{Garcia-Aspeitia2018} using different observational data sets. This demonstrate that our method can be used to constrain cosmological parameters as a complement to other standard probes. We found that $n = 7.8^{+0.9}_{-0.5}$, which shows that it deviates from a CC scenario ($n=6$) in the joint analysis at 3 $\sigma$. This result is in contrast with the one reported by \citetalias{Garcia-Aspeitia2018}, because in the present work, we found an indication of a DE that evolves as $\sim(z+1)^{-1.8}$ with an origin of extra dimensions.  Additionally, the DE expression has a singularity in $z=-1$ which represents a Big Rip in a finite value of $z$. This behavior is representative of a Phantom DE normally driven by a scalar field with a negative kinetic energy \citep{Caldwell:2003vq}.

However, our lensing analysis is in agreement with \citetalias{Garcia-Aspeitia2018} in the sense that the VBT scenario is able to source a late cosmic acceleration without an extra fluid, the five dimensional geometry  naturally produced it cosmic feature considering a brane with variable tension embedded in an extra dimension called the bulk. Thus, the VBT (\textit{chrono-brane}) scenario is a promising alternative to the problem of the accelerated expansion of the Universe without considering extra fields like the cosmological constant.

In a future study it is necessary to implement a coupling of the variable brane tension with the equation of state in order to have a more general brane-world model. Besides, as the VBT (\textit{chrono-brane}) has shown to be consistent with different observations at different scales, it should be investigated at perturbative level. These further studies could be performed using the lensing data presented here joined with other data sets, as observational Hubble data, Type Ia Supernovae, BAO, and CMB data. Moreover, our method has demonstrate to be competitive with those cosmological probes. In the coming years, high-quality imaging and spectroscopic data  will be obtained from different astronomical facilities and surveys, driven remarkable progress in strong lensing modeling of galaxy clusters \citep[e.g.][]{Bergamini2023}, and producing  the discovery of thousands of galaxy-scale
lensing system \citep[][]{Oguri2010,Collett2015}. The inclusion of strong lensing systems along with other astrophysical observations will allow a more robust analysis to comprehend the dark sectors of the Universe. 

\bmhead{Acknowledgments}
We thank the anonymous referee for thoughtful remarks and suggestions. TV, JM, MGA, AHA and VM acknowledge partial support from project ANID Vinculaci\'on Internacional FOVI220144. VM acknowledges partial support from \textit{Centro de Astrof\'{\i}sica de Valpara\'{\i}so.} M.A.G.-A. acknowledges support from c\'atedra Marcos Moshinsky (MM), Universidad Iberoamericana for support with the SNI grant. The numerical analysis was also carried out by {\it Numerical Integration for Cosmological Theory and Experiments in High-energy Astrophysics} (NICTE-HA) cluster at IBERO University, acquired through c\'atedra MM support and Graphic-Power Unit Integrated Numerical Analize (GUINA) at UCEN-Coquimbo acquired through Fondequip EQM200216. M.H.A acknowledges support from proyecto CONACYT Estancias Posdoctorales and SNI grant.

\section*{Declarations}

%Some journals require declarations to be submitted in a standardised format. Please check the Instructions for Authors of the journal to which you are submitting to see if you need to complete this section. If yes, your manuscript must contain the following sections under the heading `Declarations':

\begin{itemize}
%\item Funding
%\item Conflict of interest/Competing interests (check journal-specific guidelines for which heading to use)
%\item Ethics approval 
%\item Consent to participate
%\item Consent for publication
\item Availability of data and materials

This manuscript has no associated data or the data will not be deposited. [Authors’ comment: This is theoretical research work, and is based upon the analysis of public observational data (previously published by different authors). So no additional data are associated with this work.]

%\item Code availability 
%\item Authors' contributions
\end{itemize}

%\begin{appendices}

%\section{Section title of first appendix}\label{secA1}

%An appendix contains supplementary information that is not an essential part of the text itself but which may be helpful in providing a more comprehensive understanding of the research problem or it is information that is too cumbersome to be included in the body of the paper.

%%=============================================%%
%% For submissions to Nature Portfolio Journals %%
%% please use the heading ``Extended Data''.   %%
%%=============================================%%

%%=============================================================%%
%% Sample for another appendix section			       %%
%%=============================================================%%

%% \section{Example of another appendix section}\label{secA2}%
%% Appendices may be used for helpful, supporting or essential material that would otherwise 
%% clutter, break up or be distracting to the text. Appendices can consist of sections, figures, 
%% tables and equations etc.

%\end{appendices}

%%===========================================================================================%%
%% If you are submitting to one of the Nature Portfolio journals, using the eJP submission   %%
%% system, please include the references within the manuscript file itself. You may do this  %%
%% by copying the reference list from your .bbl file, paste it into the main manuscript .tex %%
%% file, and delete the associated \verb+\bibliography+ commands.                            %%
%%===========================================================================================%%

\bibliography{VBT_Lensing.bib}% common bib file
%% if required, the content of .bbl file can be included here once bbl is generated
%%\input sn-article.bbl

\end{document}